\documentclass[10pt]{article}
\usepackage[dvips]{graphicx}
\begin{document}
\title{Relativistic Hartree-Bogoliubov Calculation \\
of \\
Specific Heat of the Inner Crust of Neutron Stars
       \thanks{Talk presented at the 3rd JAERI Symposium on Science of Hadrons 
               under Extreme Conditions, January 2001.}}
\author{Takuya Nakano$^{1}$ and Masayuki Matsuzaki$^{2}$}
\date{}
\maketitle
\begin{center}
{\it
{\small
$^{1}$Department of Physics, Kyushu University, Fukuoka 812-8581, Japan \\
$^{2}$Department of Physics, Fukuoka University of Education, \\
Munakata, Fukuoka 811-4192, Japan
}
}
\end{center}
\vskip 1cm
\begin{abstract}
We calculate the specific heat of the inner crust of neutron stars within a 
local-density approximation to an improved relativistic Hartree-Bogoliubov 
theory. Non-uniformness of the system enhances the specific heat in 
particular at low temperatures. The degree of enhancement is similar to 
that in the spherical phase of Elgar\o y et al. We examine a schematic 
interpolation between the results of Broglia et al. adopting the Gogny 
force and ours based on the Lagrangian of the relativistic mean field model. 
\end{abstract}
\vskip 0.5cm

The inner crust of neutron stars, with densities ranging from the neutron 
drip density to that of the order of the saturation one, is a non-uniform 
system consisting of a Coulomb lattice of neutron-rich nuclei and a sea 
of free neutrons. This part conveys temperature drop due to neutrino 
emission in the core to the surface. The rate of the conveyance depends 
on the specific heat and the thickness of the inner crust. 
Since the former is sensitive to superfluidity, we study this non-uniform 
system as an application of the improved description of pairing correlation 
based on the RMF interaction, that is, a one-boson exchange interaction 
derived from the Lagrangian of the relativistic mean field (RMF) model. 
As a first step of this application, we adopt a local-density approximation 
and neglect possible non-spherical shapes of neutron-rich nuclei, in this talk. 
In this sense, this is a relativistic counterpart of Ref.\cite{nu1} and of the 
spherical part of Ref.\cite{nu2}.

The first attempt to describe superfluidity in uniform infinite nuclear matter 
adopting the RMF interaction was done in Ref.\cite{kr}; but its result was 
not very realistic in that the resulting pairing gap was too large. 
The reason can be ascribed to an unphysical behavior of the RMF interaction 
at high momenta: Repulsion is too strong. Note that repulsion also contributes 
positively to pairing gap at high momenta beyond a sign change of the 
gap function. Tanigawa and one of the present authors proposed a method to 
determine quantitatively a density-independent parameter to cut 
the momentum integration in the gap equation~\cite{tm}. Although a more 
natural way to modulate the high-momentum part smoothly was also developed 
later~\cite{mt2}, the simple method in Ref.\cite{tm} is enough for the present 
purpose since only the pairing gap at the Fermi surface at each density 
is necessary under a local-density approximation. 

\begin{figure}[htb]
 \begin{center}
  \includegraphics[width=5cm,height=5cm,clip]{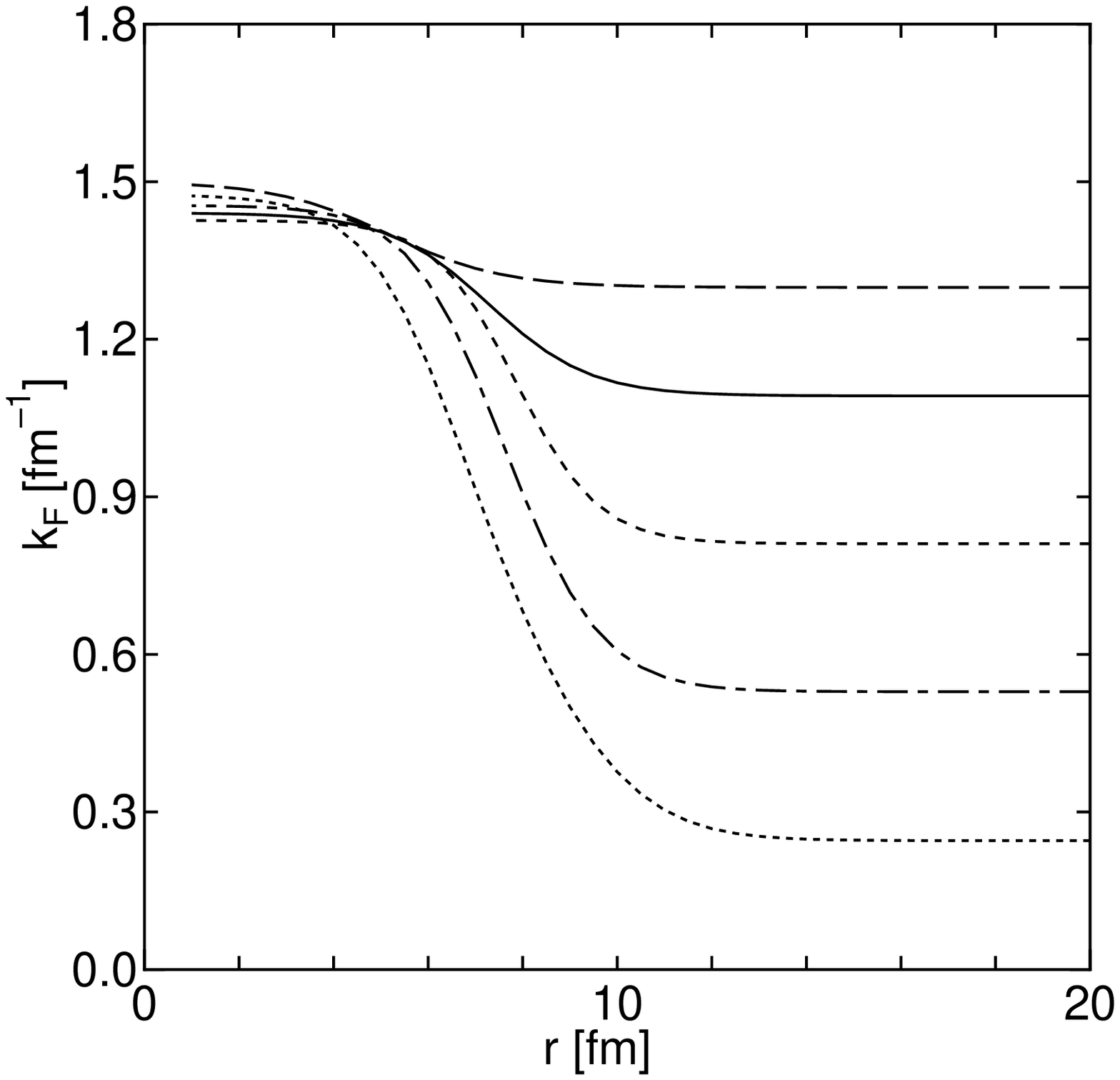}
  \includegraphics[width=5cm,clip]{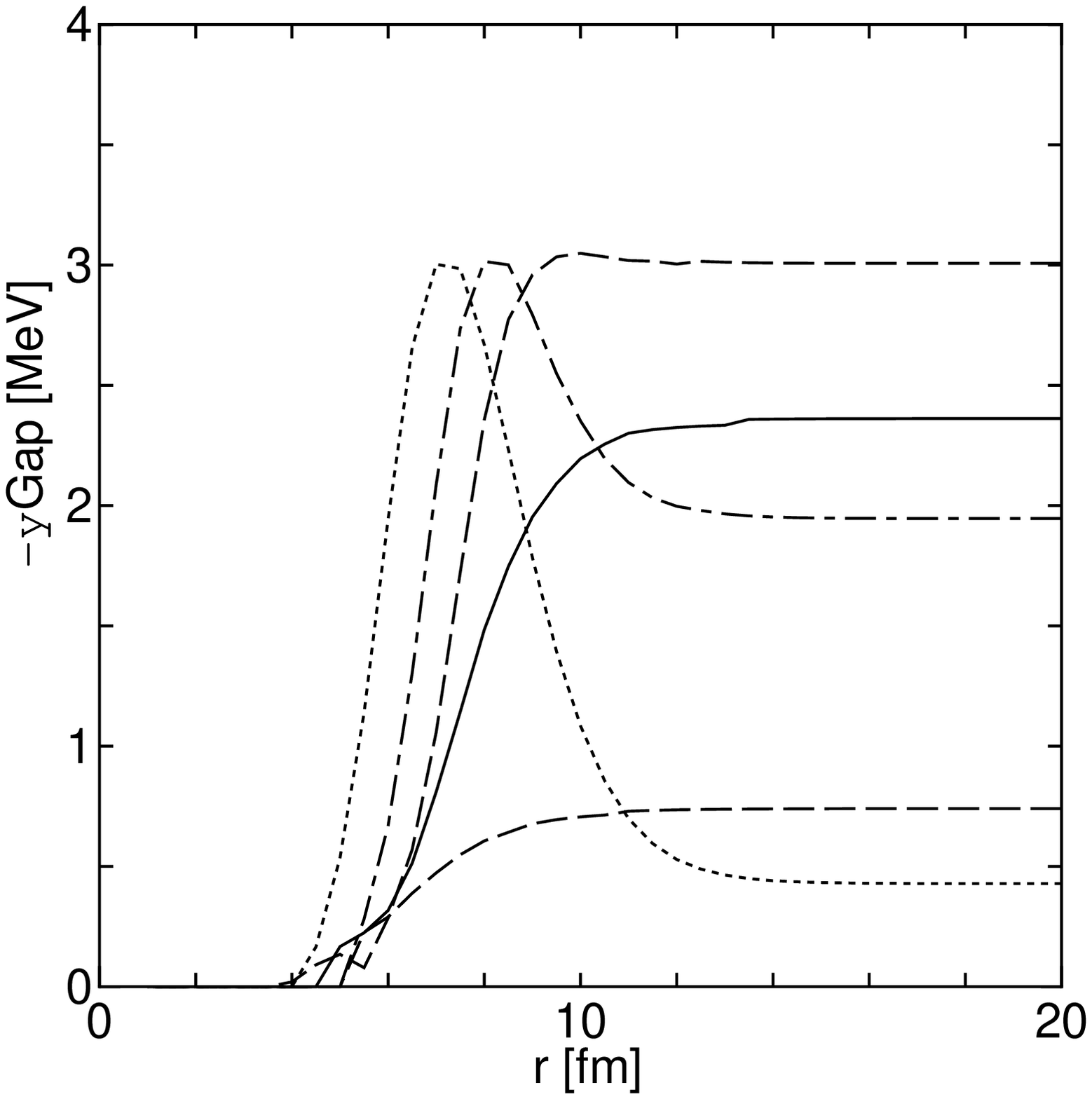}
 \end{center}
 \caption{Profile of the local Fermi momentum (left) and of the local pairing 
          gap (right) in a Wigner-Seitz cell. Five curves in the order 
          from the top to the bottom in the left panel correspond to densities 
          (a) -- (e) in Table~\ref{table1}.}
 \label{fig1}
\end{figure}

In order to describe the non-uniform system composed of a lattice of 
neutron-rich nuclei and a sea of free neutrons, we adopt the Wigner-Seitz (WS) 
approximation; each spherical cell contains at its center one nucleus 
permeated by free neutrons which dripped off from it. We parametrize the 
neutron density given by a density-dependent Hartree-Fock calculation~\cite{nv} 
in terms of a Woods-Saxon shape, 
\begin{equation}
\rho_{n}(r)=\frac{ \rho_{0,n} }{ 1+\exp [ (r-R_{n})/a_{n}] }
+\rho_{{\rm ext},n}\, ,
 \label{eq1}
\end{equation}
as done in Ref.\cite{nu1}. We perform calculations at five representative 
densities conforming to this reference; the parameters entering into 
Eq.(\ref{eq1}) and the radii of the WS cell are summarized in 
Table~\ref{table1}. Here we introduce a local-density approximation 
--- assume uniform Fermi gas of neutrons at each spatial point. 
Then the local Fermi 
momentum is calculated from the local density given by Eq.(\ref{eq1}) as 
\begin{equation}
k_{\rm F}(r)=(3\pi^2\rho_{n}(r))^{1/3}\, .
 \label{eq2}
\end{equation}
Its profile in each WS cell is shown in the left panel of Fig.\ref{fig1}.
This determines the profile of pairing gap since the gap in the 
uniform neutron matter can be calculated as a function of $k_{\rm F}$ 
by using the method of Ref.~\cite{tm}. (Note that the symmetric 
nuclear matter case was presented there.) The result is shown in the 
right panel. The local pairing gap shown there gives the local 
quasiparticle energy, 
\begin{equation}
E(r,p)=((E(p)-E(k_{\rm F}(r)))^2+\Delta(r,p)^2)^{1/2}\, ,
 \label{eq3}
\end{equation}
with $E(p)=(p^2+M^{\ast 2})^{1/2}$, $M^\ast$ being the density-dependent 
effective neutron mass.

\begin{table}[htb]
 \begin{center}
  \setlength{\tabcolsep}{6pt}
  \begin{tabular}{|c|c|c|c|c|c|c|} \hline
    &$\rho / \rho_0$ & $\rho_{0,n} [\mathrm{fm^{-3}}]$ & $\rho_{{\rm ext},n} [{\rm fm^{-3}}]$ &
   $ R_{n} [\mathrm{fm}]$ & $ a_n [\mathrm{fm}]  $& $ R_{\rm ws} [\mathrm{fm}]  $  \\ \hline 
  (a) & 0.46 & 0.114 & 0.0737 & 5.0 & 1.2 & 15.0  \\ \hline 
  (b) & 0.28 & 0.101 & 0.0436 & 7.0 & 1.1 & 20.9  \\ \hline 
  (c) & 0.12 & 0.098 & 0.0184 & 7.4 & 0.8 & 29.4  \\ \hline 
  (d) & 0.034 & 0.104 & 0.0047 & 6.8 & 0.9 & 35.5  \\ \hline 
  (e) & 0.005 & 0.108 & 0.0005 & 5.9 & 0.9 & 45.7  \\ \hline 
  \end{tabular}
 \end{center}
\caption{Parameters entering into Eq.(\ref{eq1}) and the radii of the WS cell.}
\label{table1}
\end{table}

Now we are ready to calculate the specific heat of the system. 
The specific heat is calculated from the entropy, 
\begin{equation}
S=-2\int d^{3}r\int \frac{ d^3 p }{ (2\pi)^3 }
\left\{
f(r,p)\ln f(r,p)+[1-f(r,p)]\ln[1-f(r,p)]
\right\}\, ,
 \label{eq4}
\end{equation}
as
\begin{equation}
C_{Vn}= \left. T\frac{ \partial S }{ \partial T } \right|_{V}\, .
 \label{eq5}
\end{equation}
The phase-space distribution function in Eq.(\ref{eq4}) is 
\begin{equation}
f(r,p)=\frac{1}
{ 1+\exp[ E( r,p )/T ] }\, ,
 \label{eq6}
\end{equation}
where the quasiparticle energy is given by Eq.(\ref{eq3}). 
The final expression is
\begin{equation}
C_{Vn}=\frac{1}{ \pi T V_{\rm ws}}\int_{V_{\rm ws}} dr\;r^2\;\int dp\;p^2
\frac{ E(r,p) }{ \cosh^{2}[ E(r,p)/2T ] }
\left[\frac{E(r,p)}{T}-\frac{dE}{dT} \right]\, .
 \label{eq7}
\end{equation}
The spatial integration is performed over each WS cell and the momentum 
integration is done up to $\Lambda$ = 3.60 fm$^{-1}$ determined in 
Ref.\cite{tm} to give physical pairing gaps. Note that the temperature 
dependence of the quasiparticle energy is neglected because only low 
temperatures are of interest in the present study. In addition to neutrons, 
degenerate electron gas gives a comparable contribution, 
\begin{equation}
C_{Ve}=\frac{1}{3}{k_{\rm F}^2}(m^2+k_{\rm F}^2)^{1/2}T\, ,
 \label{eq8}
\end{equation}
while that from protons is negligible. Then the total specific heat 
is given by 
\begin{equation}
C_{V{\rm tot}}=C_{Vn}+C_{Ve}\, .
 \label{eq9}
\end{equation}

\begin{figure}[htb]
 \begin{center}
  \includegraphics[width=5cm,clip]{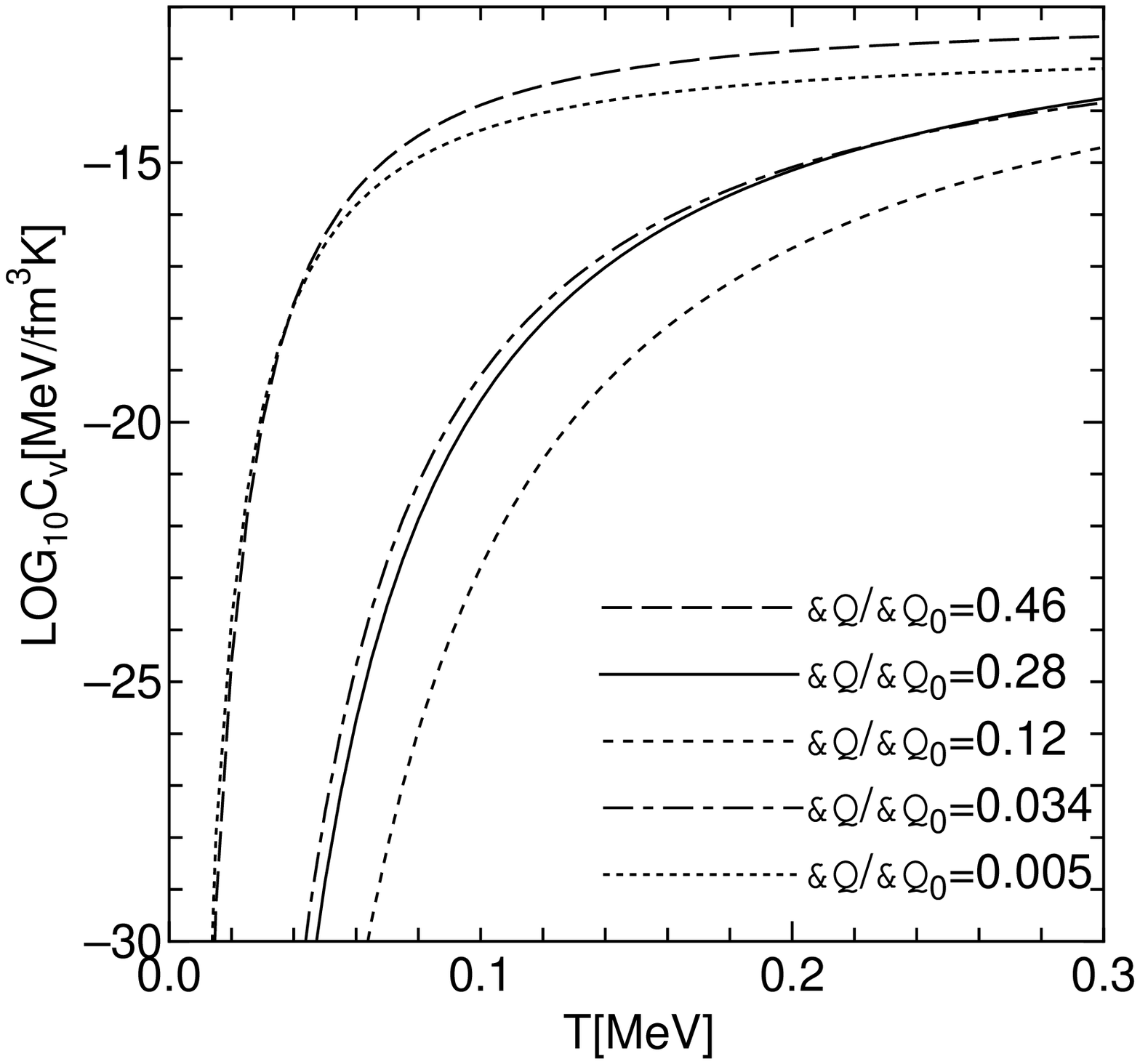}
  \includegraphics[width=5cm,clip]{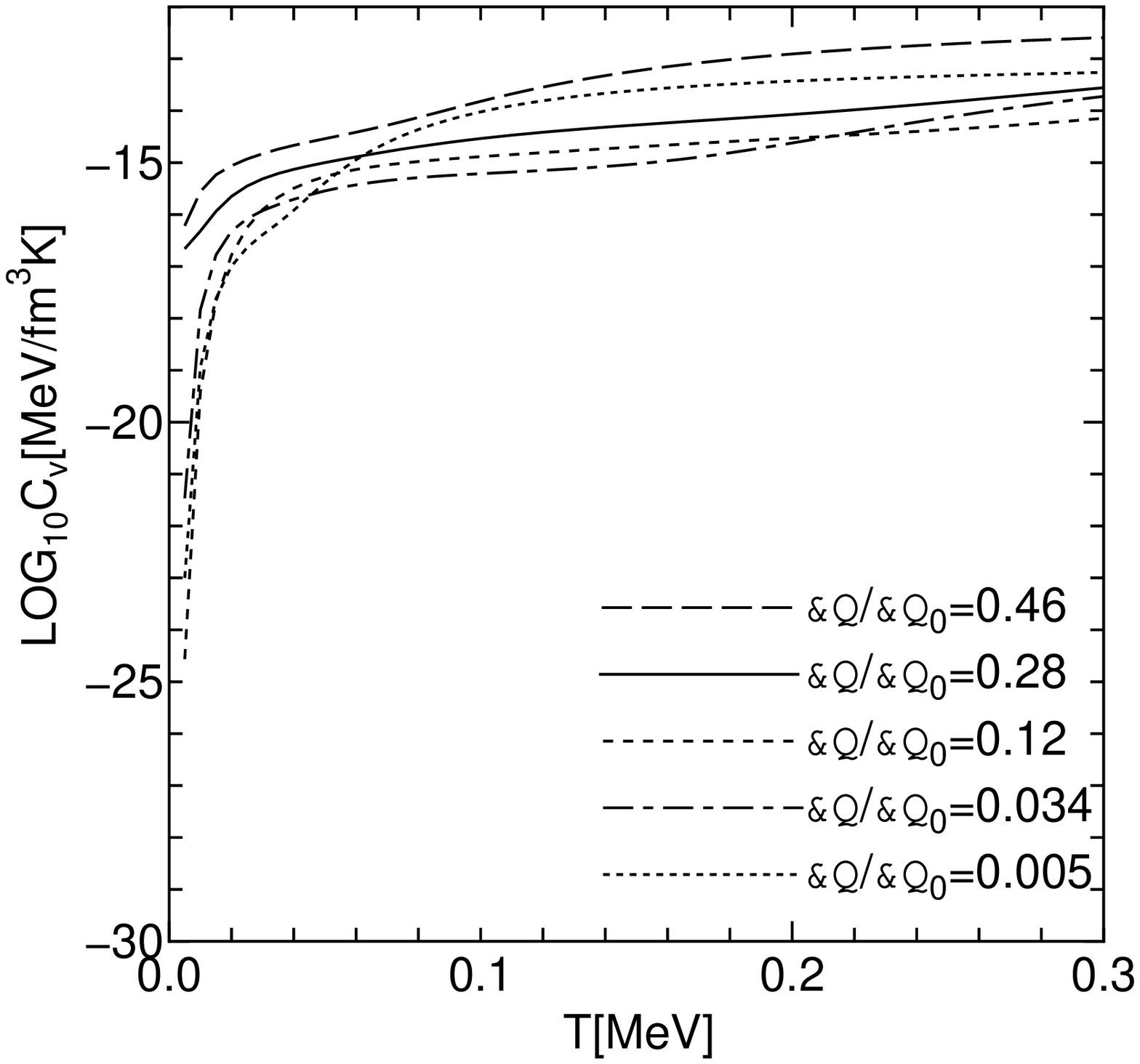}
  \includegraphics[width=5cm,clip]{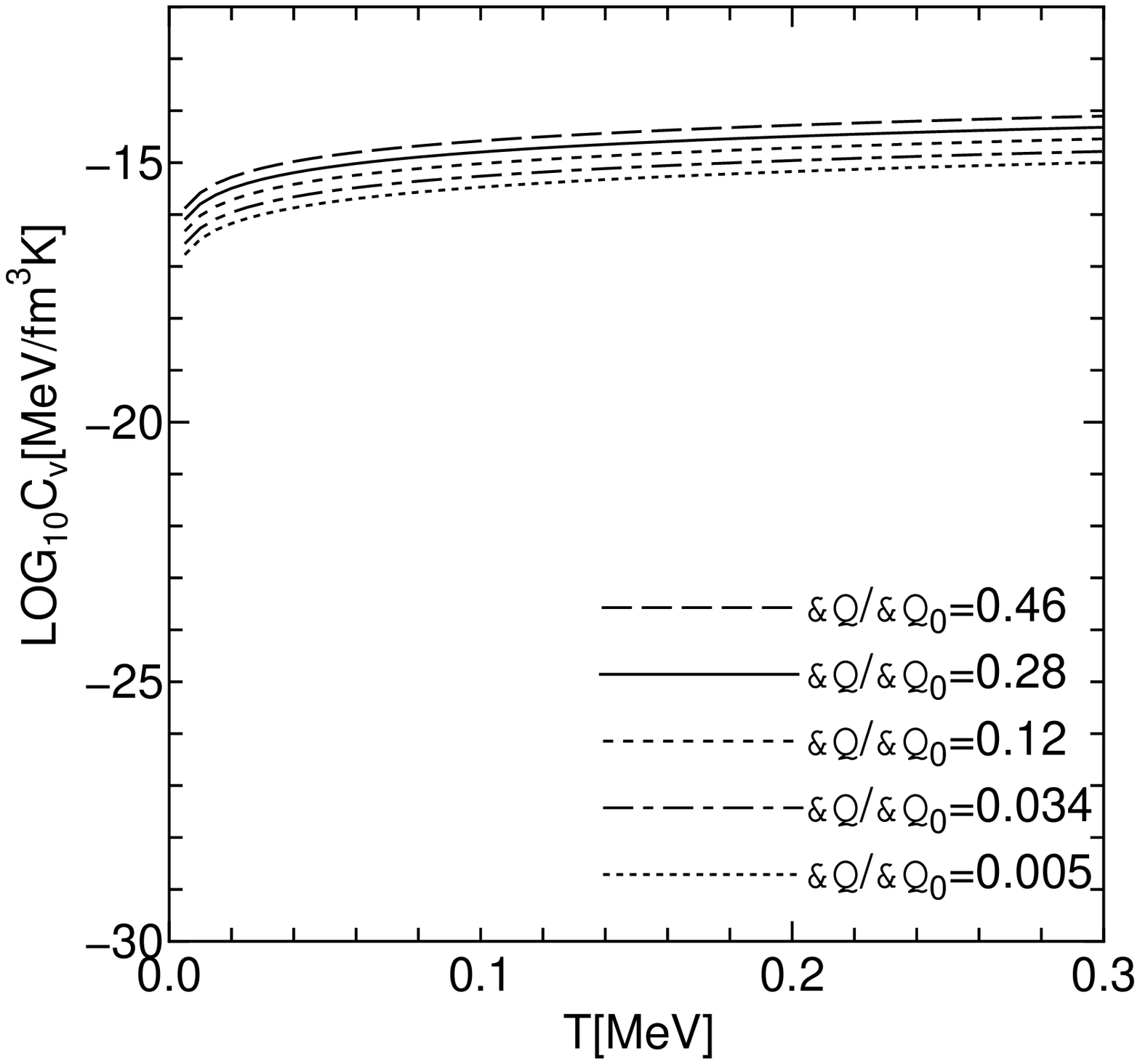}
 \end{center}
 \caption{Temperature dependence of the partial specific heat of uniform 
          neutrons (left), non-uniform neutrons (center) and electrons (right). 
         }
 \label{fig2}
\end{figure}

The effect of non-uniformness can be seen by comparing the left (showing 
$C_{Vn}$ for uniform matter) and the center (for non-uniform) panels of 
Fig.\ref{fig2}. One characteristic is that density dependence is weak 
in the non-uniform system. This is brought about by a density-dependent 
increase of $C_{Vn}$. The reason why non-uniformness increases it can 
be inferred from an expression obtained by a weak-coupling approximation 
for the uniform system~\cite{fw},
\begin{equation}
\frac{C_{Vn}}{T}\propto\left(\frac{\Delta_0}{T}\right)^{5/2}
                       \exp{\left(-\frac{\Delta_0}{T}\right)}\, ,
 \label{eq10}
\end{equation}
with $\Delta_0$ being the uniform gap. This is a decreasing function of 
$\Delta_0/T$ at low temperatures, $\Delta_0/T >$ 5/2, 
and consequently leads to
$C_{Vn}({\rm normal})>C_{Vn}({\rm super})$. Since non-uniformness  produces 
a region where the gap is small, it results in an increase of $C_{Vn}$. 
This effect is conspicuous in particular at lower temperatures because a 
small variation in $\Delta_0$ leads to a large one in $\Delta_0/T$. 
Conversely, difference in $\Delta_0$ is not important at high temperatures. 

\begin{figure}[htb]
 \begin{center}
  \includegraphics[width=5cm,clip]{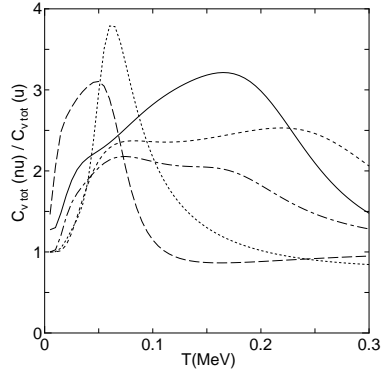}
 \end{center}
 \caption{Temperature dependence of the ratio of the total specific heat. 
          Legend of the curves is the same as in Fig.\ref{fig2}.}
 \label{fig3}
\end{figure}

Since the contribution from electrons is of the same order (the right panel 
of Fig.\ref{fig2}) as that from the non-uniform neutrons, the enhancement of
$C_{V{\rm tot}}$, 
\begin{equation}
\frac{C_{V{\rm tot}}({\rm nu})}{C_{V{\rm tot}}({\rm u})}=
\frac{C_{Vn}({\rm nu})+C_{Ve}}{C_{Vn}({\rm u})+C_{Ve}}\, ,
 \label{eq11}
\end{equation}
where ``nu" and ``u" stand for non-uniform and uniform, respectively, 
is factor 3 -- 4 as shown in Fig.\ref{fig3}.
Comparing this to Fig.3 in Ref.\cite{nu1} highlights the difference at low 
temperatures. The consideration above indicates that the difference 
originates from the region where pairing gap is small --- the interior 
region of the nucleus at the center of the WS cell. Actually, although the 
Gogny force adopted in Ref.\cite{nu1} is known to provide us with pairing 
properties similar to those given by bare interactions at low 
densities~\cite{bert}, there is a big difference at high densities --- 
superfluid phase survives up to much above the saturation density 
(see Fig.5.4 of Ref.\cite{ring}, for example). In contrast, our cutoff 
parameter in the RMF description was determined so as to reproduce the 
pairing properties given by the Bonn-B potential and accordingly pairing 
gap closes at around the saturation density. This makes the gap in the 
interior region vanish. In Ref.\cite{nu2}, the non-relativistic version 
of the Bonn-A potential was adopted, and the results for $\rho/\rho_0$ = 
0.058 and 0.176 in the spherical phase were also presented. 
Our results resemble theirs. 
Here one comment is in order on the double-counting problem of 
short-range correlations which arises when effective interactions are 
adopted in the gap equation. In principle this problem exists as pointed 
out in Ref.\cite{nu2}, but an analysis in Ref.\cite{mt2} indicates that 
it is not serious practically. 

\begin{figure}[htb]
 \begin{center}
  \includegraphics[width=5cm,clip]{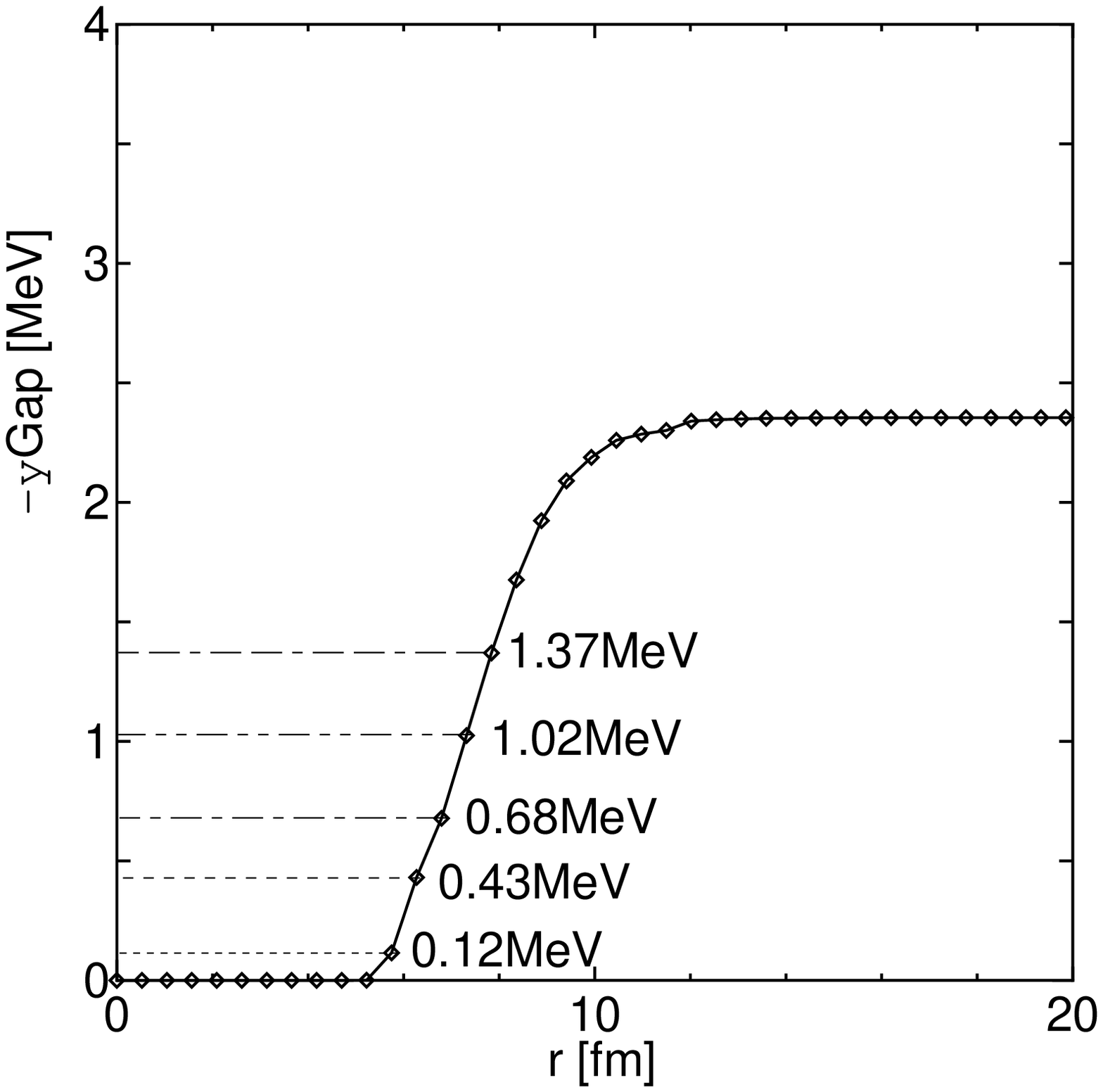}
  \includegraphics[width=5cm,height=4.95cm,clip]{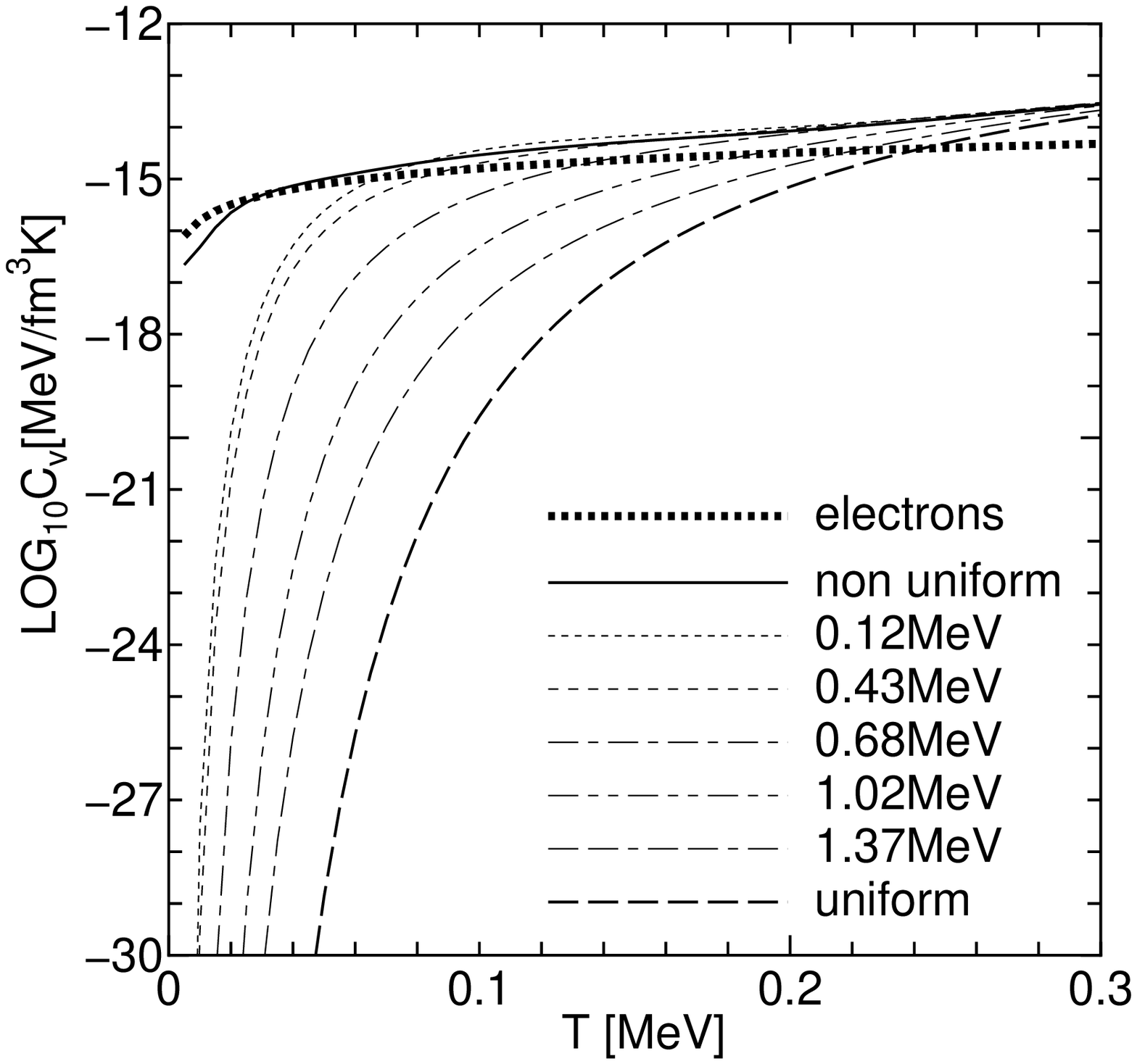}
  \includegraphics[width=5cm,clip]{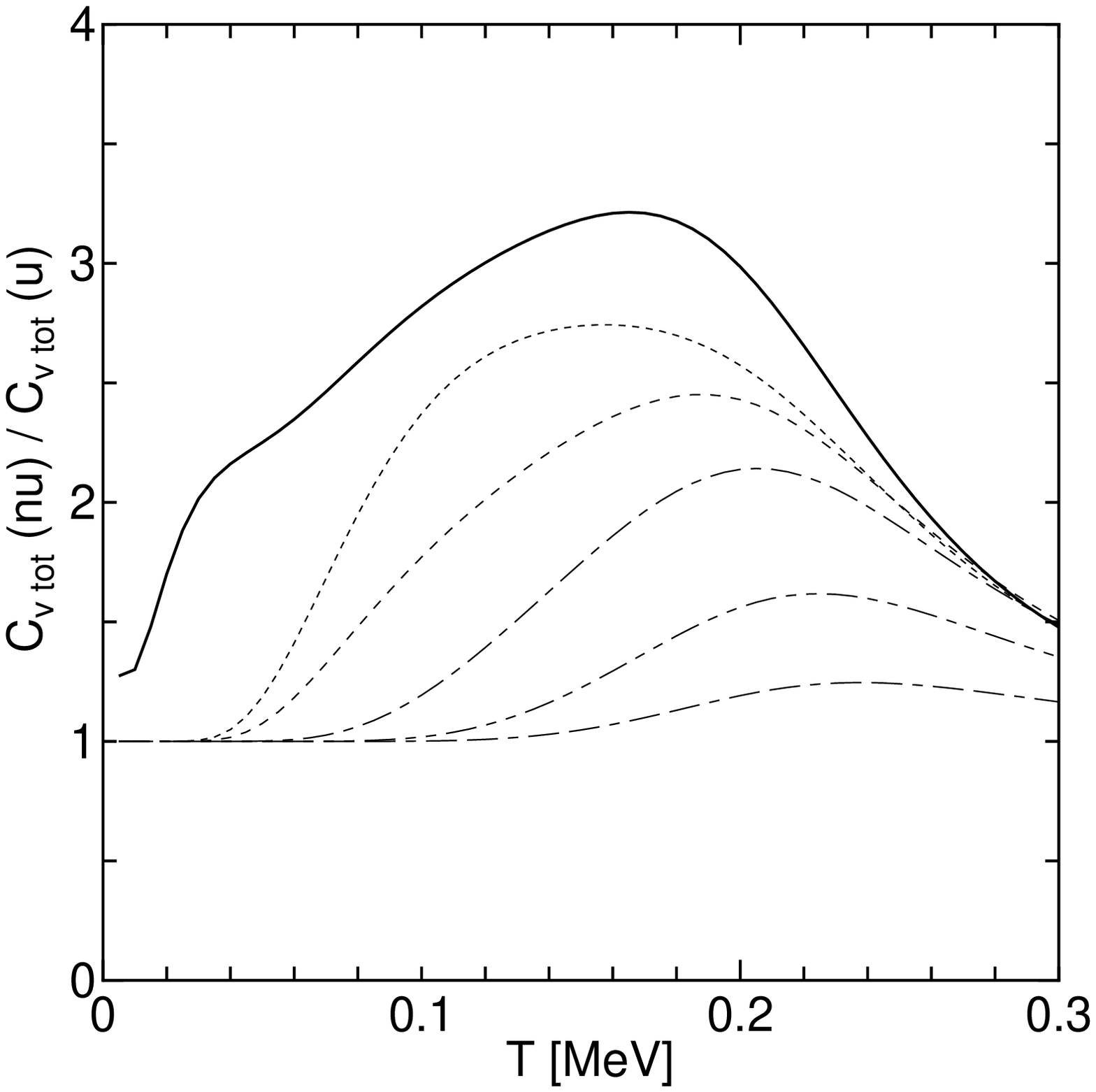}
 \end{center}
 \caption{Schematic variation of the gap profile (left), resulting temperature 
          dependence of the neutron specific heat (center) and of the ratio of 
          the total specific heat (right), at density (b).}
 \label{fig4}
\end{figure}

Since the essential difference between Ref.\cite{nu1} and ours is the gap 
in the interior region as mentioned above, we examine a schematic 
interpolation between the two models; 
we vary the gap in this region by hand as shown in the left panel 
of Fig.\ref{fig4} for density (b). The resulting specific heats are 
presented in the center panel. Clearly $C_{Vn}$ approaches to that in 
Ref.\cite{nu1} (see Fig.2 there) as the gap in the interior region is 
increased, and to that of the uniform matter by a further increase. 
Finally the total specific heats are shown in the right panel. 
This confirms that the vanishing gap in the interior region is the very 
cause of the enhancement of $C_{V{\rm tot}}$ at low temperatures. 

Beyond this first-step calculation, two kinds of refinements are in order; 
one is to take into account non-spherical shapes of lattice nuclei in the 
high-density part of the inner crust, the other is the so-called 
proximity effect beyond the local-density approximation, that is, spread 
of the pair wave function. The former was examined in Refs.\cite{nu2,nu3} 
and shown to strengthen non-uniformness, whereas the latter was shown 
to weaken it~\cite{nu4}.

\end{document}